\documentclass[prd,superscriptaddress,amsfonts,amssymb,amsmath,showpacs,twocolumn]{revtex4-2}
\usepackage{bm}
\usepackage{amsfonts}
\usepackage{latexsym}
\usepackage[latin1]{inputenc}
\usepackage{graphicx}
\usepackage{amsmath}
\usepackage{palatino}
\usepackage{mathpazo}
\usepackage{textcomp}
\usepackage{float}
\usepackage{booktabs}
\usepackage{dcolumn}
\usepackage{ragged2e}
\usepackage{hyperref}
\hypersetup{colorlinks,citecolor=green}
\hypersetup{colorlinks=true,linkcolor=blue,filecolor=blue,    urlcolor=magenta}
\usepackage{amsmath}
\usepackage{xcolor}
\usepackage[caption=false]{subfig}
\usepackage{commath}
\def\jnl@style{\it}
\def\aaref@jnl#1{{\jnl@style#1}}

\def\aaref@jnl#1{{\jnl@style#1}}

\def\aj{\aaref@jnl{AJ}}                   
\def\apj{\aaref@jnl{ApJ}}                 
\def\apjl{\aaref@jnl{ApJ}}                
\def\apjs{\aaref@jnl{ApJS}}               
\def\apss{\aaref@jnl{Ap\&SS}}             
\def\aap{\aaref@jnl{A\&A}}                
\def\aapr{\aaref@jnl{A\&A~Rev.}}          
\def\aaps{\aaref@jnl{A\&AS}}              
\def\mnras{\aaref@jnl{Mon.~Not.~Roy.~Astron.~Soc.}}             
\def\prd{\aaref@jnl{Phys.~Rev.~D}}        
\def\plb{\aaref@jnl{Phys.~Lett.~B}}        
\def\prc{\aaref@jnl{Phys.~Rev.~C}}  
\def\prl{\aaref@jnl{Phys.~Rev.~Lett.}}    
\def\qjras{\aaref@jnl{QJRAS}}             
\def\skytel{\aaref@jnl{S\&T}}             
\def\ssr{\aaref@jnl{Space~Sci.~Rev.}}     
\def\zap{\aaref@jnl{ZAp}}                 
\def\nat{\aaref@jnl{Nature}}              
\def\aplett{\aaref@jnl{Astrophys.~Lett.}} 
\def\apspr{\aaref@jnl{Astrophys.~Space~Phys.~Res.}} 
\def\physrep{\aaref@jnl{Phys.~Rep.}}      
\def\physscr{\aaref@jnl{Phys.~Scr}}       
\def\commat{\aaref@jnl{Comm.~Math.~Phys.}}              
\def\science{\aaref@jnl{Science}}               
\def\cqg{\aaref@jnl{Classical Quant.~Grav.}}            
\def\jpcs{\aaref@jnl{JPCS}}                                     
\def\ijmpd{\aaref@jnl{Int.~J.~Mod.~Phys.~D}}                    
\def\grg{\aaref@jnl{Gen.~Relat.~Gravit.}}               
\def\rpp{\aaref@jnl{Rep.~Prog.~Phys.}}          
\def\npa{\aaref@jnl{Nucl.~Phys.~A}}        
\def\lrr{\aaref@jnl{Living Rev.~Rel.}}                   
\def\jcap{\aaref@jnl{J.~Cosmology Astropart.~Phys.}}    
\def\rmp{\aaref@jnl{Rev.~Mod.~Phys.}}   
\def\epjc{\aaref@jnl{Eur.~Phys.~J.~C}}

\allowdisplaybreaks[1]

\begin{document}

\color{black}       

\title{The Mass-Radius relation for Quark Stars in Energy-Momentum Squared Gravity }

\author{Takol Tangphati}  
\email{takoltang@gmail.com}
\affiliation{
Theoretical and Computational Physics Group, \\
Theoretical and Computational Science Center (TaCS), Faculty of Science, \\
King Mongkut's University of Technology Thonburi, 126 Prachauthid Rd., Bangkok 10140, Thailand
}

\author{Indrani Karar } 
\email{indrani.karar08@gmail.com}
\affiliation{Department of Registrar, University of Kalyani, Nadia, West Bengal, India}

\author{Ayan Banerjee} 
\email{ayanbanerjeemath@gmail.com}
\affiliation{Astrophysics and Cosmology Research Unit, School of Mathematics, Statistics and Computer Science, University of KwaZulu--Natal, Private Bag X54001, Durban 4000, South Africa}

\author{Anirudh Pradhan}
\email{pradhan.anirudh@gmail.com}
\affiliation{Centre for Cosmology, Astrophysics and Space Science, GLA University, Mathura-281 406, Uttar Pradesh, India}


\date{\today}

\begin{abstract}

We study the structure of quark stars (QSs) adopting homogeneously confined matter inside the star with a 3-flavor neutral charge and a fixed strange quark mass $m_s$. We explore the internal structure, and the physical properties of specific classes of QSs in the recently proposed energy-momentum squared gravity (EMSG).
Also, we obtain the mass-radius $(M-R)$ and mass-central energy density $(M-\rho_c)$ relations for QS using the QCD motivated EoS. The maximum mass for QSs in EMSG is investigated depending on the presence and absence of the free parameter $\alpha$. Furthermore, the stability of stars is determined by the condition
$\frac{d M}{d \rho_c}>0$.  We observe that consideration of the EMSG  has specific contributions to the structure of QSs.

\end{abstract}

\maketitle

\section{Introduction}

Of course, Modified gravity theories (MGT) are a new paradigm of modern physics for explaining the major problems in Einstein's theory of General Relativity (GR). Assuming this point of view, MGT became very popular due to its ability to
provide an alternative framework instead of searching for new material ingredients. Thus, one of the expected outcomes and impact of MGT is to address the phenomenology of gravity at galactic, extragalactic, and cosmological scales \cite{Nojiri:2006ri,Capozziello:2007ec}.

However, it is not easy to construct gravity theories fulfilling these requirements.  Aiming these descriptions, a new gravity model has been proposed in \cite{Katirci:2013okf} where the
Einstein-Hilbert action is replaced by an arbitrary function of Ricci scalar $R$ and of the norm of energy-momentum tensor i.e., $f(R,T_{\mu \nu}T^{\mu \nu})$. Such a generalization of GR includes new type of contributions to the right-hand side of the Einstein field equations. Recently, the authors of Ref. \cite{Roshan:2016mbt} discussed a particular example of this type of generalizations in the form
$f(R,\textbf{T}^2) = \mathcal{R}+\alpha \textbf{T}^2$, known as energy-momentum squared gravity (EMSG). So, the matter fields are not conserved in this theory. This idea has been more generalized by adding higher-order terms of the form $f(R,\textbf{T}^2) = \mathcal{R}+\alpha (\textbf{T}^2)^n$, dubbed as energy-momentum powered gravity (EMPG) \cite{Akarsu:2017ohj,Board:2017ign}. This corresponds to EMSG in the case $n = 1$. In Ref. \cite{Akarsu:2019ygx}, the authors proposed a logarithmic generalization, known as energy-momentum log gravity (EMLG), 
where the term $\ln \left(\lambda T_{\mu \nu}T^{\mu \nu}\right)$ was used for an extension of the standard
$\Lambda$CDM model to study viable cosmologies. Such matter-geometry coupling theories are promising in constructing novel cosmological models (see \cite{Akarsu:2018drb,Faria:2019ejh,Barbar:2019rfn,Akarsu:2020vii} for a review). Besides cosmological solutions, astrophysical solutions have also been explored, so they do not conflict with GR experimental results. For instance, the possibility of an existing wormhole solution was developed recently in \cite{Moraes:2017dbs,Sharif:2021ptz}.

From a genuine mathematical point of view, the strong gravity regime is another way to check the viability of these theories. Regarding this 
stellar evolution can be considered a suitable test-beds for EMSG. 
Besides, that compact objects (i.e., neutron stars and white dwarfs) are formed by the gravitational collapse
of intermediate-mass stars is an excellent laboratory for studying highly dense matter in extreme conditions. But to date, the composition and properties of compact objects are not well understood theoretically. 

On the other hand, the discovery of pulsars with their masses and radii poses considerable constraints on the high-density region of the nuclear matter equation of state (EoS) and, consequently, the interior composition of neutron stars (NSs) \cite{Lattimer:2015nhk}. They were commonly regarded as neutron stars (NSs) until Gell-Mann proposed the idea of quark stars (QSs) and Zweig \cite{GellMann:1964nj,Zweig:1964jf}. Nowadays, it is generally believed that NSs was the best candidate in the universe where quark matter could be found. Central density in the inner core of NSs may reach values several times higher than the nuclear saturation density,
where a phase transition from nuclear to quark matter occurs \cite{Alcock:1996qb,Madsen:1998qb,Glendenning:1992kd}. This quark matter may be the true ground state of hadronic matter \cite{Witten:1984rs,Bodmer:1971we}. Based on the 'quark matter' hypothesis the heaviest stars are interpreted as QSs. Various authors have been working in this area for a long time, mostly constraining the QS
mass-radius using different quark matter EoSs \cite{Astashenok:2014dja,Sedaghat:2022fue,Parisi:2020qfs,Baym:2017whm,Astashenok:2014dja}. 

Addressing the problem of exotic relativistic stars in modified gravity, we examine the existence of QSs
with interacting quark EoS in the framework of EMSG.  In the recent past, several works have been devoted to study compact objects in the framework of EMSG, see Refs. \cite{Akarsu:2018zxl,Nari:2018aqs,Carvalho:2019ert,Sharif:2021axd}. In particular, Singh \textit{et al} \cite{Singh:2020bdv}
have shown the existence of a stable, compact star in EMSG theory constraining the value of coupling constant
with the recent constraints on the mass-radius relation. 

In the present study, we consider the interacting EoS as our primary input to obtain a configuration for QS and satisfy the current observational data on the $(M-R)$ relations. The paper is organized as follows: In Section \ref{sec2}, we present the field equations for EMSG  and derive the equations of motion. The structure of QS is developed by assuming the case of the spherically symmetric metric of these equations and obtaining the modified Tolman-Oppenheimer-Volkoff (TOV) equations. In Section \ref{sec3}, we present an overview of
a QCD motivated EoS. Numerically we solved the modified TOV equations with appropriate boundary conditions in Section \ref{sec4}. We found that the presence of $\alpha$ (EMSG coupling constant) can lead to larger masses
for QSs from the predictions of GR. Finally, we summarize the results in the conclusion section \ref{sec5}.


\section{Field equations in energy-momentum EMSG gravity (EMSG)}\label{sec2}

We start with the action of the EMSG gravity \cite{Roshan:2016mbt} ( we set $G = 1$ and $c = 1$ in the action and in the rest of this section)
\begin{equation}
S= {\frac{1}{2\kappa}} \int \left( ~\mathcal{R}+\alpha \textbf{T}^2 \right) \sqrt{-g}~d^4x + \int \mathcal{L}_m  \sqrt{-g}~d^4x, \label{e1}
\end{equation}
where $\textbf{T}^2=T_{\mu \nu}T^{\mu \nu}$, $T_{\mu \nu}$ is the energy-momentum tensor,  $g$ is the determinant of the
metric, $R$ is the scalar curvature and  $\kappa = 8 \pi$.  The $\mathcal{L}_m$ denotes the Lagrangian density corresponding
to the matter source and define by
\begin{equation}\label{emt}
T_{\mu \nu} = -{2 \over \sqrt{-g}} {\delta (\sqrt{-g} \mathcal{L}_m) \over \delta g^{\mu \nu}}=\mathcal{L}_m~g_{\mu \nu}-2 {\partial \mathcal{L}_m \over \partial g^{\mu \nu}},
\end{equation}
which depends only on the metric
tensor components $g_{\mu \nu}$, and not on its derivatives. 
For the sake of simplicity the stress-energy tensor is usually assumed to be a perfect fluid form
\begin{eqnarray}
T_{\mu \nu} = (\rho+P)u_\mu u_\nu+Pg_{\mu \nu}, \label{eq3}
\end{eqnarray}
where $\rho$ is the energy density, $P$ is the (isotropic) pressure, $u^\mu$ is the (timelike) 4-velocity of the 
fluid and $g_{\mu \nu}$ is the metric tensor. Varying the action (\ref{e1}) with respect to the metric, one obtains the following gravitational field equations
\begin{eqnarray}\label{eq4}
\mathcal{G}_{\mu \nu} = 8\pi T_{\mu \nu}+8\pi \alpha\left(g_{\mu \nu} T_{\beta \gamma}T^{\beta \gamma}-2\Theta_{\mu \nu} \right), \label{eq4}
\end{eqnarray}
 where $\mathcal{R}_{\mu \nu}-{1 \over 2}\mathcal{R} g_{\mu \nu}$ is the Einstein tensor. The new tensor $\Theta_{\mu \nu}$ is defined as
 \begin{eqnarray}\label{eq5}
\Theta_{\mu \nu} &=& T^{\beta \gamma} {\delta T_{\beta \gamma} \over \delta g^{\mu \nu}}+T_{\beta \gamma} {\delta T^{\beta \gamma} \over \delta g^{\mu \nu}} \nonumber \\
&=& -2\mathcal{L}_m (T_{\mu \nu}-{1\over 2}g_{\mu \nu}T)-TT_{\mu \nu}+2T^\gamma_\mu T_{\nu \gamma}- \nonumber\\
&& 4T^{\beta \gamma} {\partial^2 \mathcal{L}_m \over \partial g^{\mu \nu} \partial g^{\beta \gamma}},
\end{eqnarray}
where $T=g^{\mu \nu}T_{\mu \nu}$, the trace of EMT. As it is well know that there is no unique
definition of the matter Lagrangian; one could choose $\mathcal{L}_m = P$ or $\mathcal{L}_m = -\rho$.
In the current paper we consider $\mathcal{L}_m=P$ and the covariant divergence of Eq. (\ref{emt}) reads
\begin{eqnarray}
\nabla^\mu T_{\mu \nu}=-\alpha g_{\mu \nu} \nabla^\mu (T_{\beta \gamma} T^{\beta \gamma})+2\alpha \nabla^\mu \Theta_{\mu \nu}. \label{eq6}
\end{eqnarray}
Note that the local/covariant energy-momentum conservation $\nabla^\mu T_{\mu \nu} $ is not identically zero,
for $\alpha \neq 0$. 

Now, substituting Eq. \eqref{eq3} in Eq. \eqref{eq5}, and then using the resultant equation in Eq. \eqref{eq4},
the field equation becomes  
\begin{eqnarray}
&&  \mathcal{R}_{\mu \nu}-{1 \over 2}\mathcal{R} g_{\mu \nu} = 8\pi \rho \bigg[\left(1+{P \over \rho} \right)u_\mu u_\nu+{P \over \rho} g_{\mu \nu} \bigg]+8\pi \alpha \rho^2 \nonumber \\
&&  \bigg[2\left(1+{4P \over \rho} + {3P^3 \over \rho^2} \right) u_\mu u_\nu + \left(1+{3P^2 \over \rho^2} \right)g_{\mu \nu} \bigg]. \label{eq7}
\end{eqnarray}

To derive the modified field equations that describe the internal structure of a star in EMSG, 
we choose the spherically symmetric line element in the form
\begin{eqnarray}
\hspace{-0.5cm} ds^2 &=& - e^{2\nu} dt^2+e^{2\lambda} dr^2+r^2(d\theta^2+\sin^2 \theta ~d\phi^2), \label{eq8}
\end{eqnarray}
where $\nu(r)$ and $\lambda(r)$ are functions of the radial coordinate, $r$. Using the metric given in Eq. (\ref{eq8}) in Eq. (\ref{eq7}), the $(tt)$ and $(rr)$ components are (see Ref. \cite{Akarsu:2018zxl})
\begin{eqnarray}
{e^{-2\lambda} \over r^2} \left(2r \lambda'-1 \right)+{1 \over r^2}= \rho_{\mbox{eff}}(r), \label{eq9}\\
 {e^{-2\lambda} \over r^2} \left(2r \nu' +1 \right)-{1 \over r^2}  = P_{\mbox{eff}}(r), \label{eq10}
\end{eqnarray}
where prime denotes derivative with respect to $r$. The effective density and pressure are given by 
\begin{eqnarray}
\rho_{\mbox{eff}}(r)&=& 8\pi \rho +8\pi \alpha \rho^2\left(1+{8P \over \rho}+ {3P^2 \over \rho^2}\right), \nonumber \\
P_{\mbox{eff}}(r) &=&  8\pi P +8\pi \alpha \rho^2\left(1+{3P^2 \over \rho^2}\right). \nonumber 
\end{eqnarray}

By representing the metric tensor coefficient $e^{-2\lambda}$ as
\begin{eqnarray}
e^{-2\lambda} = 1-{2m(r) \over r}, \label{eq11}
\end{eqnarray}
where $m(r)$ is the total mass inside the sphere of radius $r$.
The other metric function, $\nu(r)$, is related to the pressure via
\begin{eqnarray}
\frac{d \nu}{dr} &=& -\left[\rho \left(1+{P \over \rho}\right) \left\{1+2\alpha \rho \left( 1+{3P \over \rho}\right) \right\} \right]^{-1} \nonumber \\
&& \Big[(1+6\alpha P) P'(r)+2\alpha \rho \rho'(r) \Big], \label{e12}
\end{eqnarray}
 Finally, solving the continuity equations (\ref{eq9})-(\ref{e12}), the modified TOV equations take the following convenient form \cite{Akarsu:2017ohj} 
\begin{eqnarray}
m'(r)=4 \pi  \rho  r^2 \left[1+\alpha  \rho  \left(\frac{3 P^2}{\rho ^2}+\frac{8 P}{\rho }+1\right)\right], \label{eq13}
\end{eqnarray}
and
\begin{eqnarray}
P'(r) &=& -\frac{m \rho }{r^2} \left(1+\frac{P}{\rho }\right) \left( 1-\frac{2 m}{r}\right)^{-1} \bigg[1+\frac{4 \pi  P r^3}{m}+ \alpha \nonumber \\
&& \frac{4 \text{$\pi $r}^3 \rho ^2}{m} \left(\frac{3 P^2}{\rho ^2}+1\right) \bigg] \bigg[1+2 \alpha  \rho  \left(1+\frac{3 P}{\rho }\right) \bigg] \nonumber \\
&& \bigg[1+2 \alpha  \rho  \left({d\rho \over dP}+\frac{3 P}{\rho }\right) \bigg]^{-1}. \label{eq14}
\end{eqnarray}
Naturally, by setting the parameter $\alpha =0$, the
standard TOV equation is recovered. Similar to the case in GR,  the modified TOV equations are solved numerically with
a specific EoS relating to the pressure and energy density of the fluid. 

\section{Quark matter equation of state} \label{sec3}

The basic idea relies on a strange star consisting of strange quark matter (SQM), composed of almost equal numbers of $u$, $d$, and $s$ quarks, and a small number of electrons
to attain the charge neutrality. As a result, SQM may be the
the true ground state of matter \cite{Itoh:1970uw,Bodmer:1971we}, because its energy per baryon could be less than that of the most stable atomic nucleus, such as $^{56}$Fe and $^{62}$Ni. The possible existence of quark matter in compact stars is an issue that has attracted considerable attention for a long time. 

In this work, we use a model presented in \cite{Flores:2017kte}, which is based on
the homogeneously confined matter inside the star with a 3-flavor neutral charge and a fixed strange quark mass $m_s$.
 Within this model, one can describe this phase using the simple thermodynamic Bag model EoS \cite{Alford:2004pf}
with $\mathcal{O}$ $(m_s^4)$ corrections. Such EoS can be expressed via explicit expressions for energy density
($\rho$) and pressure ($p$) \cite{Becerra-Vergara:2019uzm} 
\begin{align} \label{Prad1}
  p =&\ \dfrac{1}{3}\left(\rho-4B\right)-\dfrac{m_{s}^{2}}{3\pi}\sqrt{\dfrac{\rho-B}{a_4}}  \nonumber  \\
  &+ \dfrac{m_{s}^{4}}{12\pi^{2}}\left[1-\dfrac{1}{a_4}+3\ln\left(\dfrac{8\pi}{3m_{s}^{2}}\sqrt{\dfrac{\rho-B}{a_4}}\right)\right],
\end{align}
where the interacting parameter $a_4$ provides relatively tight
constraints on the quark matter EoS (see, e.g., \cite{Becerra-Vergara:2019uzm, Banerjee:2020dad, Panotopoulos:2021sbf}). In \cite{Alford:2004pf}, authors showed the existence of $M \approx 2 M_{\odot}$ hybrid stars for $a_4 \approx 0.7$. Using the same formalism, several authors have studied the influence of $a_4$ on mass-radius $(M-R)$ relations of QSs
\cite{Tangphati:2021tcy, Tangphati:2021mvu}. The strange quark mass $m_{s}$ has been taken $100\, {\rm MeV}$ \cite{Beringer:2012} and the Bag constant $B$ lies in the range of  $57\leq B \leq 92\, \rm MeV/fm^3$ \cite{Burgio:2018mcr, Blaschke:2018mqw}. 

Now, using the EoS we are in a position to close the system of equations. To ensure regularity of interior spacetime,  we specify the following initial conditions at the center of the star
\begin{align}\label{BC1}
   \rho(0) &= \rho_c,  ~~~~    \text{and} ~~~~  m(0) = 0 ,
\end{align}
where $\rho_c$ is the central energy density,  and we integrate outwards up to the pressure vanishes. The radius of the star $r_s$ is identified by the condition
\begin{eqnarray}\label{NSR}
p(r_s)=0,  
 \end{eqnarray} 
where $r_{\rm s}$ is the radius of the star. For the exterior solution, we assume a Schwarzschild solution as in GR, and hence the continuity of the metric at the surface imposes a boundary 
condition which yields
\begin{eqnarray}
e^{-2\lambda} = 1-{2M \over r_s}, \label{e11}
\end{eqnarray}
where $m(r_{\rm sur}) \equiv M$ is the total mass at the surface of the star.


\section{Numerical results}\label{sec4}

The modified TOV equations (\ref{eq13})-(\ref{eq14}) combined with a quark matter EoS does not have an analytical
solution. Thus, we numerically solve the set of equations starting from the origin $r = 0$ to the surface at $r = r_s$ where the pressure vanishes. Here, we have four parameters;  the MIT bag constant $B$ in MeV/fm$^3$, the strange quark mass $m_s$ in MeV, the interacting parameter $a_4$, and the coupling constant $\alpha$ in cm$^3$/erg. In our calculations the mass of the stars is measured in solar masses, $M_{\odot}$ and the radius of the stars in ${\rm km}$.

Using the method described above, we present the numerical results based on the variation of three sets of parameters; 
(i) the variation of negative values of $\alpha$ in the range of $-3.0 \times 10^{-38} \leq \alpha \leq 0$ cm$^3$/erg (ii) the variation of positive values of $\alpha$ in the range of $0 \leq \alpha \leq 3.0 \times 10^{-38}$ cm$^3$/erg (iii) the variation of $a_4$ in the range of $(0, 1)$.
In order to compare the structure of the stars in EMSG and standard GR, we presented the solutions of $M-R$, $M-\rho_c$ and compactness $(M-M/R)$ in Figs. \ref{figVaryAlpha}-\ref{figGammaVaryA4}, respectively. Moreover, obtained results are enlisted in Tables \ref{tableVaryAlpha}-\ref{tableVaryA4}.

\subsection{The mass-radius relations}

\begin{table}[h]
    \caption{Properties of QSs in EMSG: The maximum masses and corresponding their radii have been tabulated for specific central energy density $\rho_c$.  Results are given for different values $\alpha \in [-3.0 \times 10^{-38}, 0]$ with other parameters are same as of Fig. \ref{figVaryAlpha}. The compactness $M/R$ is measured as a dimensionless quantity. Typical value of $\alpha =0$ leads to the GR solution.}
    \begin{ruledtabular}
    \begin{tabular}{ccccc}
    $\alpha \times 10^{-38}$  & $M$ & $R$ & $\rho_c$ & $M/R$\\
     cm$^3$/erg & $M_{\odot}$ &   km & MeV/fm$^3$ & \\
    \hline
        0.0 (GR) &  1.782  & 9.814 & 1379 & 0.2691 \\
        -0.5 &  1.786  & 9.827 & 1435 & 0.2694 \\
        -1.0 &  1.790  & 9.865 & 1463 & 0.2689 \\
        -1.5 &  1.794  & 9.875 & 1547 & 0.2692 \\
        -2.0 &  1.797  & 9.899 & 1632 & 0.2690 \\
        -2.5 &  1.799  & 9.926 & 1745 & 0.2687 \\
        -3.0 &  1.802  & 9.950 & 1970 & 0.2684 \\
    \end{tabular}
    \end{ruledtabular}
    \label{tableVaryAlpha}
\end{table}

\begin{figure}[h]
    \centering
    \includegraphics[width= 7.5cm]{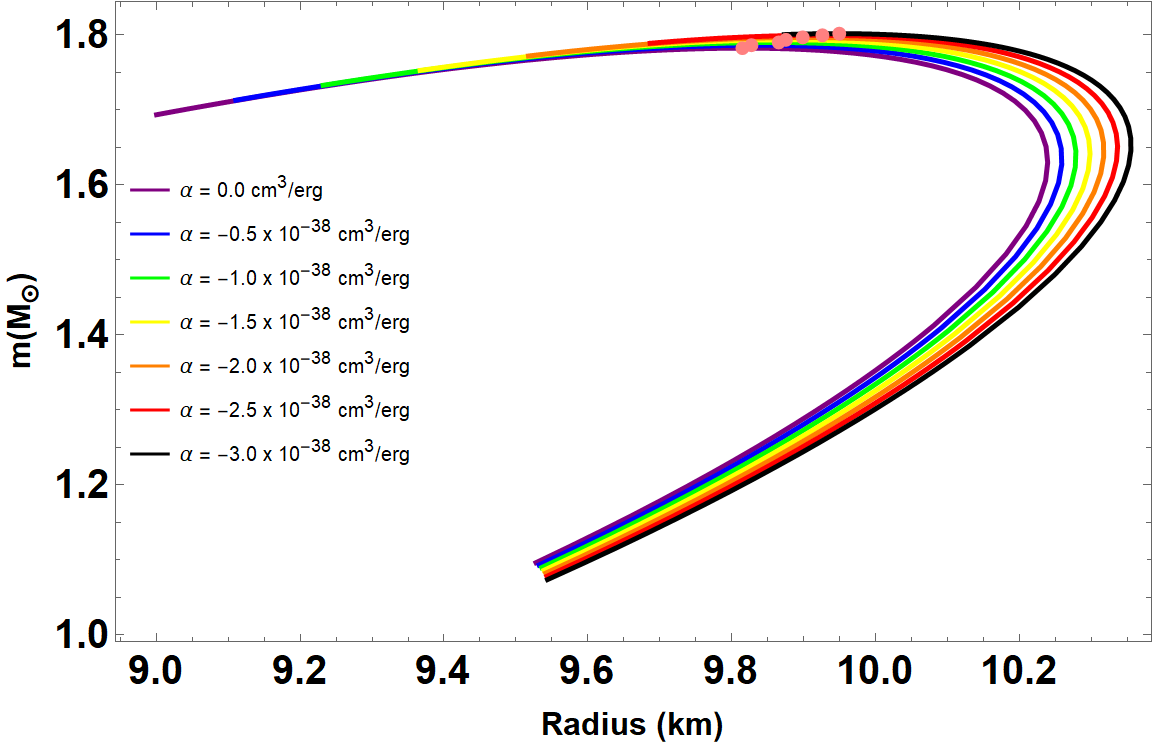}
    \includegraphics[width= 7.5cm]{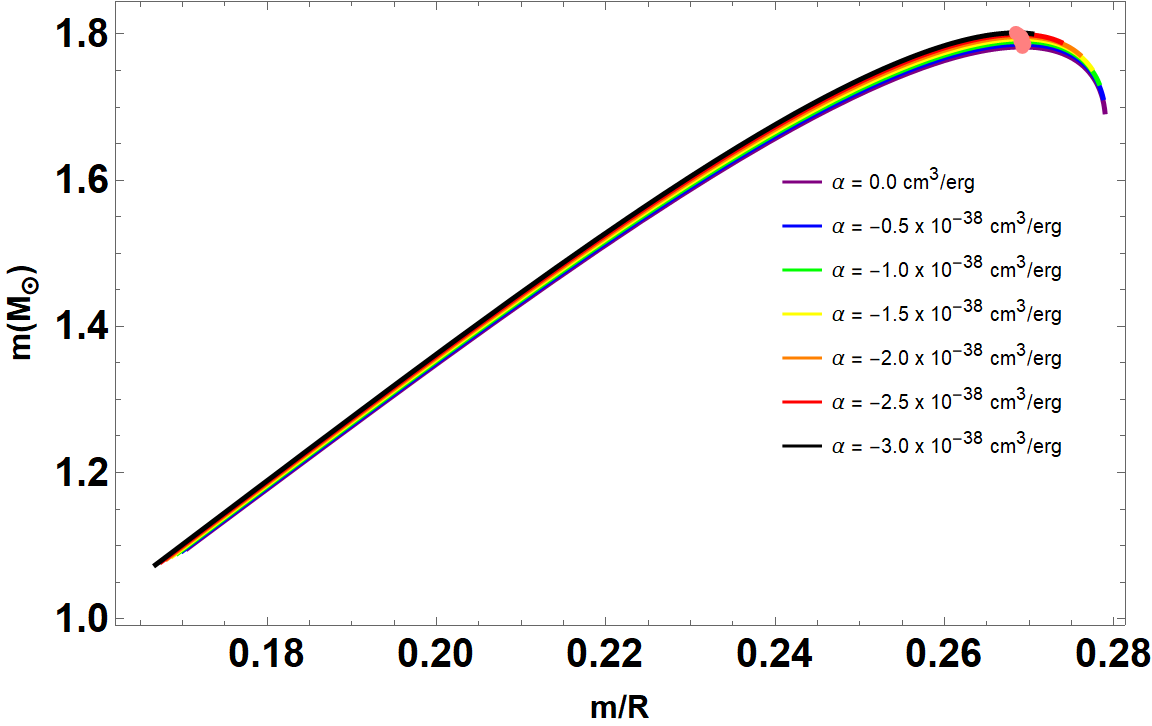}
    \caption{The mass-radius $(M-R$) relation and the compactness $(M-M/R)$ of QSs have been plotted for $B = 70$ MeV/fm$^3$, $a_4 = 0.7$, $m_s = 50.0$ MeV and $\alpha \in [-3.0,0.0] \times 10^{-38}$ cm$^3$/erg.}
    \label{figVaryAlpha}
\end{figure}

\begin{figure}[h]
    \centering
    \includegraphics[width = 7.5 cm]{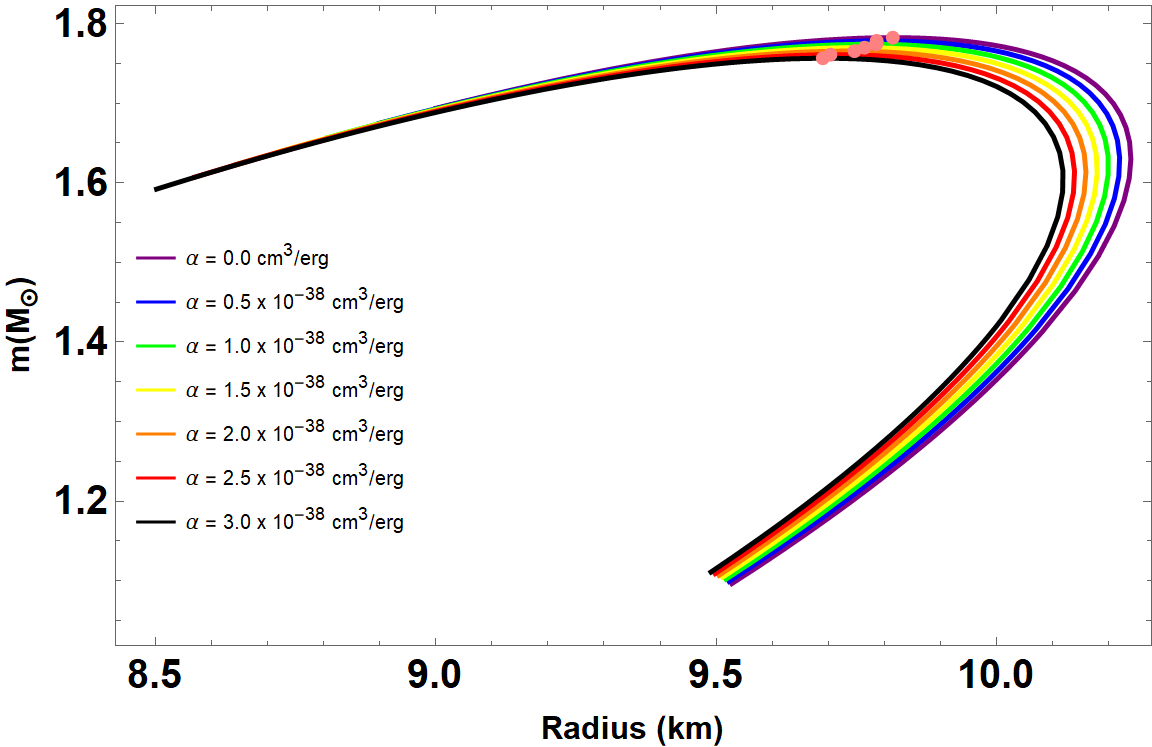}
    \includegraphics[width = 7.5 cm]{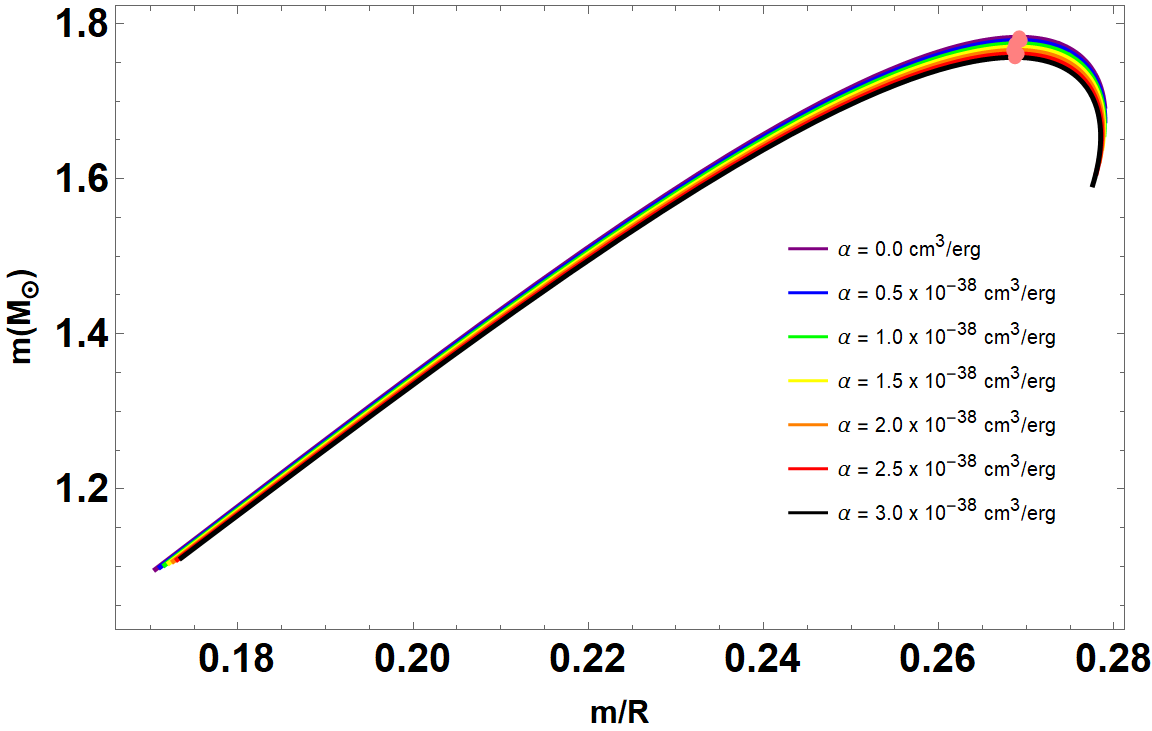}
    \caption{The mass-radius $(M-R$) relation and the compactness $(M-M/R)$ of QSs have been plotted for $B = 70$ MeV/fm$^3$, $a_4 = 0.7$, $m_s = 50.0$ MeV and $\alpha \in [0.0, 3.0] \times 10^{-38}$ cm$^3$/erg.}
    \label{figVaryAlpha2_positive}
\end{figure}

For a given quark matter EoS (\ref{Prad1}),  we plot mass-radius $(M-R)$ and compactness ($M-M/R$) of QSs in Figs. \ref{figVaryAlpha} and  \ref{figVaryAlpha2_positive} for 
negative and positive values of $\alpha$. We show results for different values $\alpha \in [-3.0 \times 10^{-38}, 3.0 \times 10^{-38}]$ cm$^3$/erg 
with other parameters are $B = 70$ MeV/fm$^3$, $a_4 = 0.7$ and $m_s = 50$ MeV. The purple solid curves correspond to GR solution ($\alpha =0$). According to Figs. \ref{figVaryAlpha} and  \ref{figVaryAlpha2_positive}, $\alpha$ has an impact on the maximum mass and its radius of the star in EMSG gravity. We have seen that the star attains its maximum mass for decreasing values of $\alpha$, when 
$\alpha <0$. The maximal mass is 1.802 $M_{\odot}$ for $\alpha = - 3.0 \times 10^{-38}$ cm$^3$/erg, see Table \ref{tableVaryAlpha}. Interestingly this situation is opposite for $\alpha >0$ i.e.,  the maximum mass decreases for increasing values of $\alpha $. It is seen from the Fig. \ref{figVaryAlpha2_positive}  and Table \ref{tableVaryAlpha2_pos} that EMSG leads to smaller masses for QSs compared with GR for $\alpha >0$. For an example the maximum mass reaches 1.778 $M_{\odot}$ for $\alpha = 0.5 \times 10^{-38}$ cm$^3$/erg in EMSG, whereas GR takes  1.782 $M_{\odot}$ for $\alpha =0$.
However the maximum mass difference compared to GR solution is not too significant, see Table \ref{tableVaryAlpha} and  \ref{tableVaryAlpha2_pos} for details. From our 
present discussion we see that the maximum gravitational mass never reach 2$M_{\odot}$ limit, because our numerical solution fails to integrate the structure   equations when  $\alpha < - 3.0 \times 10^{-38}$ cm$^3$/erg.  More specifically, there is no stable  QS solution when $\alpha < - 3.0 \times 10^{-38}$ cm$^3$/erg and the decreasing values of $\alpha$ does not guarantees that $dP_{\mbox{eff}}/dr < 0$ everywhere throughout the star. Moreover, the influence of $\alpha$ is reflecting on the compactness of QSs in EMSG. The results have been shown in the lower panels of Figs. \ref{figVaryAlpha} and  \ref{figVaryAlpha2_positive}, respectively. For $\alpha < 0$, we obtain the values of compactness $(M/R)$ in the Table \ref{tableVaryAlpha}. Table \ref{tableVaryAlpha} shows the maximum compactness decreases with decreasing values of $\alpha < 0$ and lies within the range of  $0.2684 < M/R < 0.2694$. On the other hand, Table \ref{tableVaryAlpha2_pos} reflects the $M/R$ relations of the star with respect to the maximum mass for $\alpha > 0$. We see that the maximum compactness decreases with increasing the values of $\alpha$. This situation is completely opposite to the previous case. Considering the positive values of $\alpha$, the maximum compactness lies in the range of $0.2687 < M/R < 0.2693$. To see this fact we can say that compactness stays nearly the same for both cases.

\begin{table}[h]
    \caption{Properties of QSs in EMSG: The maximum masses and corresponding their radii have been tabulated for specific central energy density $\rho_c$.  Results are given for different values $\alpha \in [0, 3.0 \times 10^{-38}]$ with other parameters are same as of Fig. \ref{figVaryAlpha}. The compactness $M/R$ is measured as a dimensionless quantity. Typical value of $\alpha =0$ leads to the GR solution.}
    \begin{ruledtabular}
    \begin{tabular}{ccccc}
    $\alpha \times 10^{-38}$  & $M$ & $R$ & $\rho_c$ & $M/R$\\
     cm$^3$/erg & $M_{\odot}$ &   km & MeV/fm$^3$ & \\
    \hline
        0.0 (GR) &  1.782  & 9.814 & 1379 & 0.2691 \\
        0.5 &  1.778  & 9.786 & 1350 & 0.2693 \\
        1.0 &  1.774  & 9.785 & 1295 & 0.2687 \\
        1.5 &  1.770  & 9.764 & 1266 & 0.2686 \\
        2.0 &  1.765  & 9.747 & 1238 & 0.2685 \\
        2.5 &  1.761  & 9.704 & 1238 & 0.2690 \\
        3.0 &  1.756  & 9.690 & 1210 & 0.2687 \\
    \end{tabular}
    \end{ruledtabular}
    \label{tableVaryAlpha2_pos}
\end{table}

\begin{table}[h]
    \caption{Properties of QSs in EMSG gravity: The maximum masses and corresponding their radii have been tabulated for specific central energy density $\rho_c$.  Results are given for different values $a_4 \in [0.1, 0.9]$ and a fixed value of $\alpha = 3.0 \times 10^{-38}$ cm$^3$/erg. Other parameters are same as of Fig. \ref{figVaryAlpha}. }
    \begin{ruledtabular}
    \begin{tabular}{ccccc}
    $a_4 $  & $M$ & $R$ & $\rho_c$ & $M/R$\\
     & $M_{\odot}$ &   km & MeV/fm$^3$ & \\
    \hline
        0.1 & 1.718 & 9.507 & 1252 & 0.2679 \\
        0.3 & 1.744 & 9.631 & 1224 & 0.2684 \\
        0.5 & 1.752 & 9.674 & 1210 & 0.2685 \\
        0.7 & 1.756 & 9.690 & 1210 & 0.2687 \\
        0.9 & 1.759 & 9.700 & 1210 & 0.2688 \\
    \end{tabular}
    \end{ruledtabular}
    \label{tableVaryA4}
\end{table}

\begin{figure}[h]
    \centering
    \includegraphics[width = 7.5 cm]{MRPlotMarkedVaryPosAlpha.png}
    \includegraphics[width = 7.5 cm]{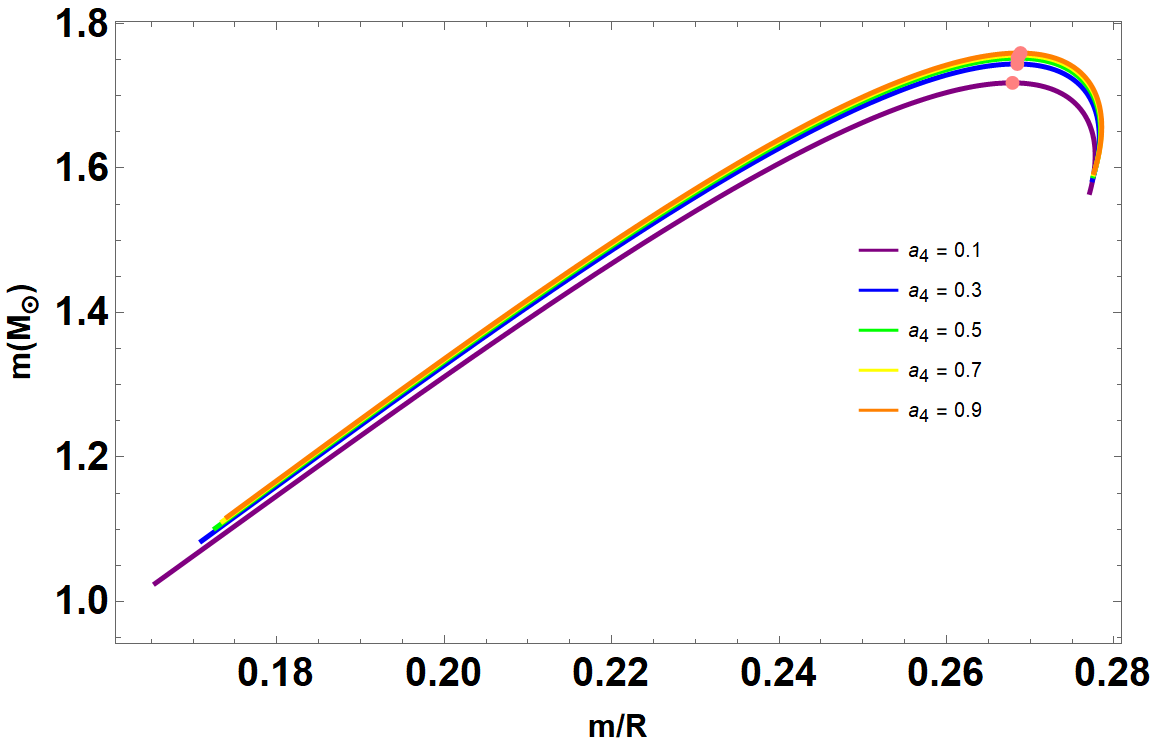}
    \caption{The mass-radius $(M-R$) relation and the compactness $(M-M/R)$ of QSs have been plotted for $B = 70$ MeV/fm$^3$, $m_s = 50$ MeV, $\alpha = 3.0 \times 10^{-38}$ cm$^3$/erg and $a_4 \in [0.1, 0.9]$.}
    \label{figVaryA4}
\end{figure}

Put in the above way, we plot Fig. \ref{figVaryA4} with the set of parameters; $B = 70$ MeV/fm$^3$, $m_s = 50$ MeV, $\alpha = 3.0 \times 10^{-38}$ cm$^3$/erg and $a_4 \in [0.1, 0.9]$. We choose $\alpha$ in such a way to get the maximum effect on the star model, which is already recorded in the previous section. The trend of $(M-R)$ and ($M-M/R$) relations are almost similar to Figs. \ref{figVaryAlpha} and  \ref{figVaryAlpha2_positive}, respectively. Table \ref{tableVaryA4} summarizes the main aspects of the model used in this work and we recorded the maximum mass is 1.759$M_{\odot}$ at $a_4 = 0.9$. We see that the gravitational mass increases with increasing $a_4$.  Considering the EMSG one can obtain the maximum compactness $\frac{M}{R} \sim 0.2688$ at $a_4 = 0.9$, which is consistent with the $M-R$ relations as of Fig. \ref{figVaryA4}.

\subsection{The stability criterion and the adiabatic indices}

\begin{figure}[h]
    \centering
    \includegraphics[width= 7.5cm]{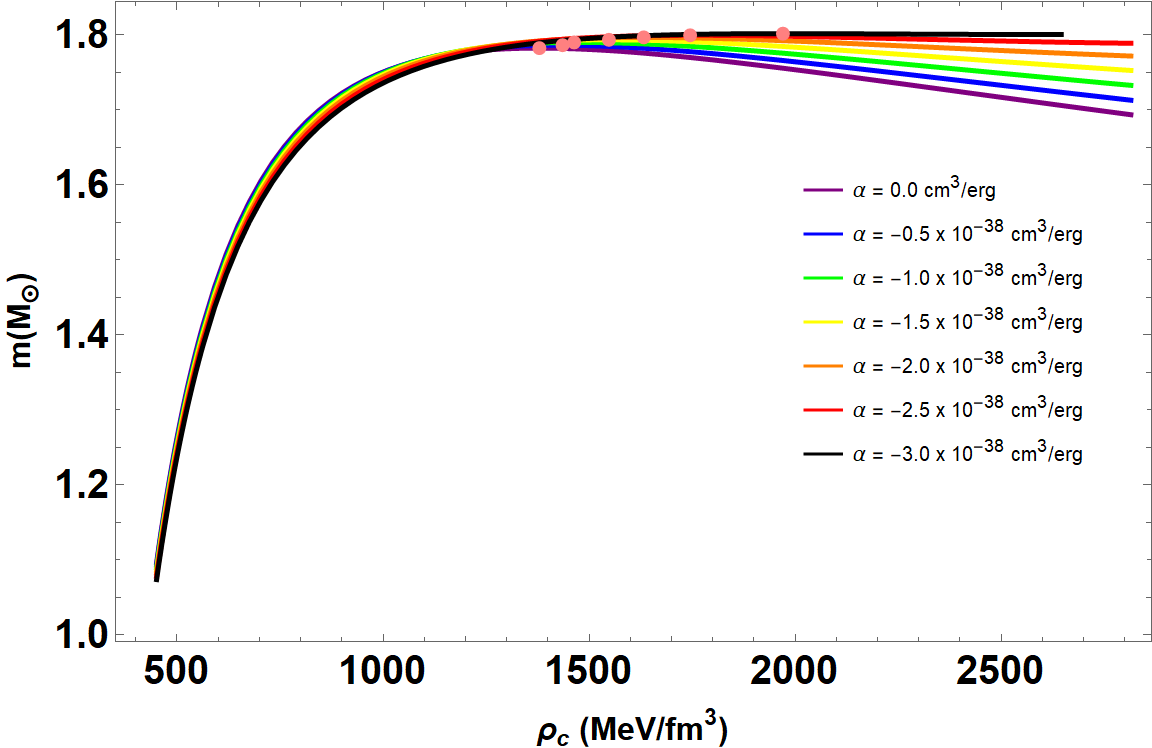}
    \includegraphics[width=7.5cm]{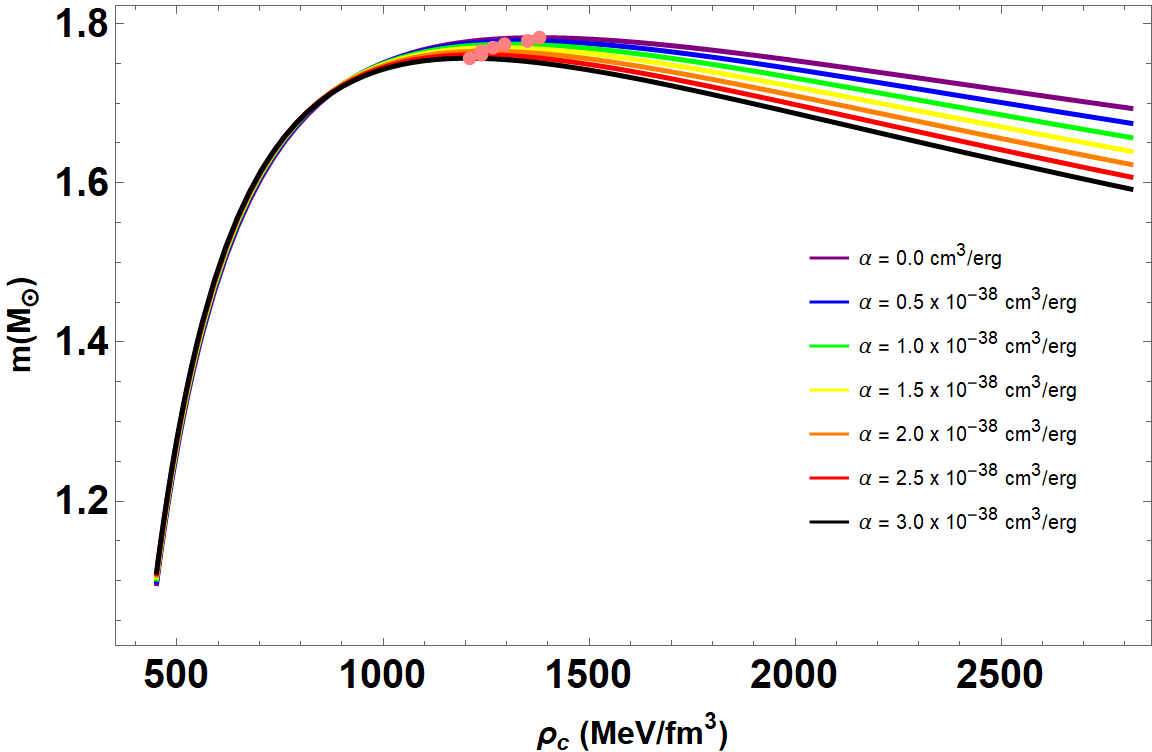}
    \includegraphics[width= 7.7cm]{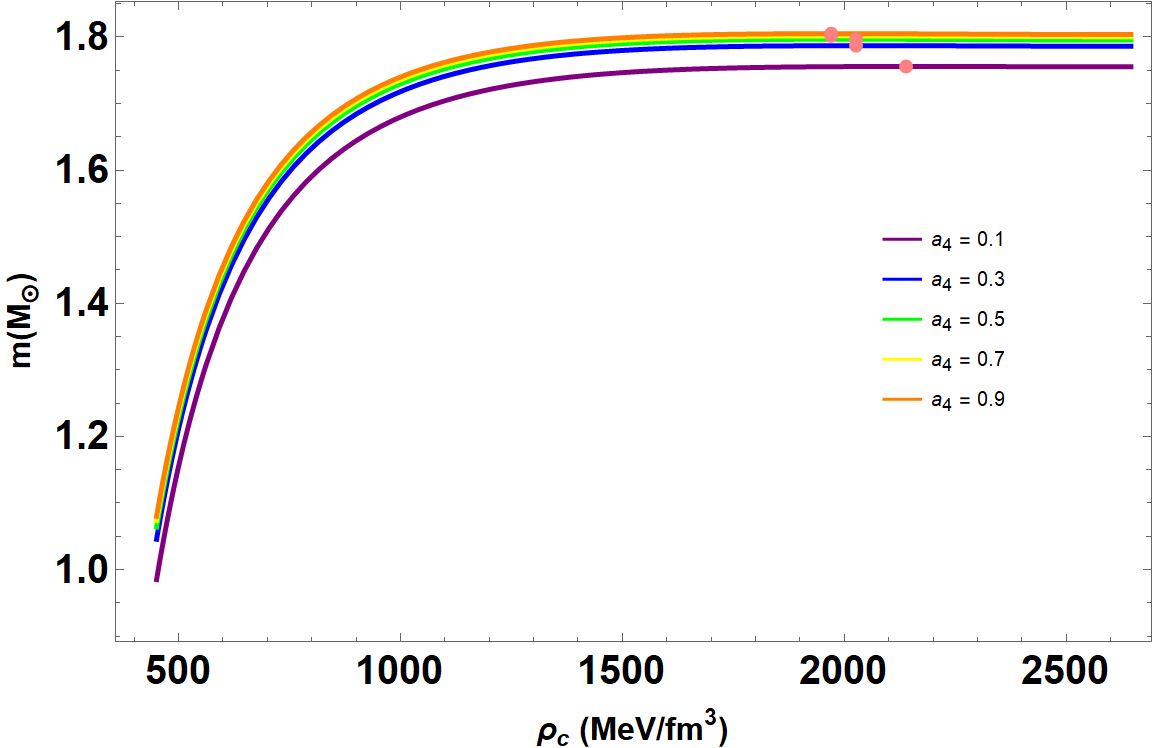}
    \caption{Gravitational mass $M$ is plotted against the central energy density $\rho_c$. Two different values of  $\alpha \in [-3.0,3.0] \times 10^{-38}$ cm$^3$/erg (split into two panels) and $a_4 \in [0.1,0.9]$ are considered ( see the captions of Figs. \ref{figVaryAlpha}, \ref{figVaryAlpha2_positive} and \ref{figVaryA4}). The circle on each $(M-\rho_c)$ curves correspond to the maximum-mass configuration.}
    \label{fig_ME}
\end{figure}

The stability of the stellar model is one of the most important criteria. To address this question, one can analyze the relation between total mass $M$ and the central
energy density $\rho_c$ at the turning point where $dM/d\rho_c = 0$. From the definition of \textit{static stability
criterion} \cite{harrison,ZN}, which reads
\begin{eqnarray}
    &&\frac{d M}{d \rho_c} < 0 ~~~ \rightarrow \text{unstable configuration}, \\
    &&\frac{d M}{d \rho_c} > 0 ~~~ \rightarrow \text{stable configuration},
    \label{criterion_M_rho_c}
\end{eqnarray}
to be satisfied by all stellar configurations. The criteria mentioned above represent a boundary that separates the stable configuration region from the unstable one at the point
$(M_{\text{max}}, R_{M_{\text{max}}})$. But, it is a necessary condition but not sufficient. In Fig. \ref{fig_ME}, we plot QSs as a function of the stellar mass $M$ and the central energy density $\rho_c$, which grows with decreasing values of $\alpha$ and increasing values of $a_4$ in agreement with the $M-R$ profiles given in  Figs.  \ref{figVaryAlpha},  \ref{figVaryAlpha2_positive} and  \ref{figVaryA4}, respectively.  The pink circles on each $M-\rho_c$ curve correspond to the maximum mass of stellar configuration.

Another way to check the stability of QS constructed in EMSG gravity we need to study the adiabatic index $\gamma$ of these stars. The corresponding study was first performed by Chandrasekhar \cite{Chandrasekhar}
based on the variational method to determine the dynamical stability of spherical configurations
of a perfect fluid. The adiabatic index $\gamma$ can be written as follows,
\begin{eqnarray}\label{adi}
    \gamma \equiv \left(1+\frac{\rho}{P}\right)\left(\frac{dP}{d\rho}\right)_S,
\end{eqnarray}
where  $dP/ d\rho$  is the speed of sound and the subscript $S$ indicates the derivation at constant entropy. In general, Eq. (\ref{adi}) is a dimensionless quantity.

\begin{figure}
    \centering
    \includegraphics[width = 7.5 cm]{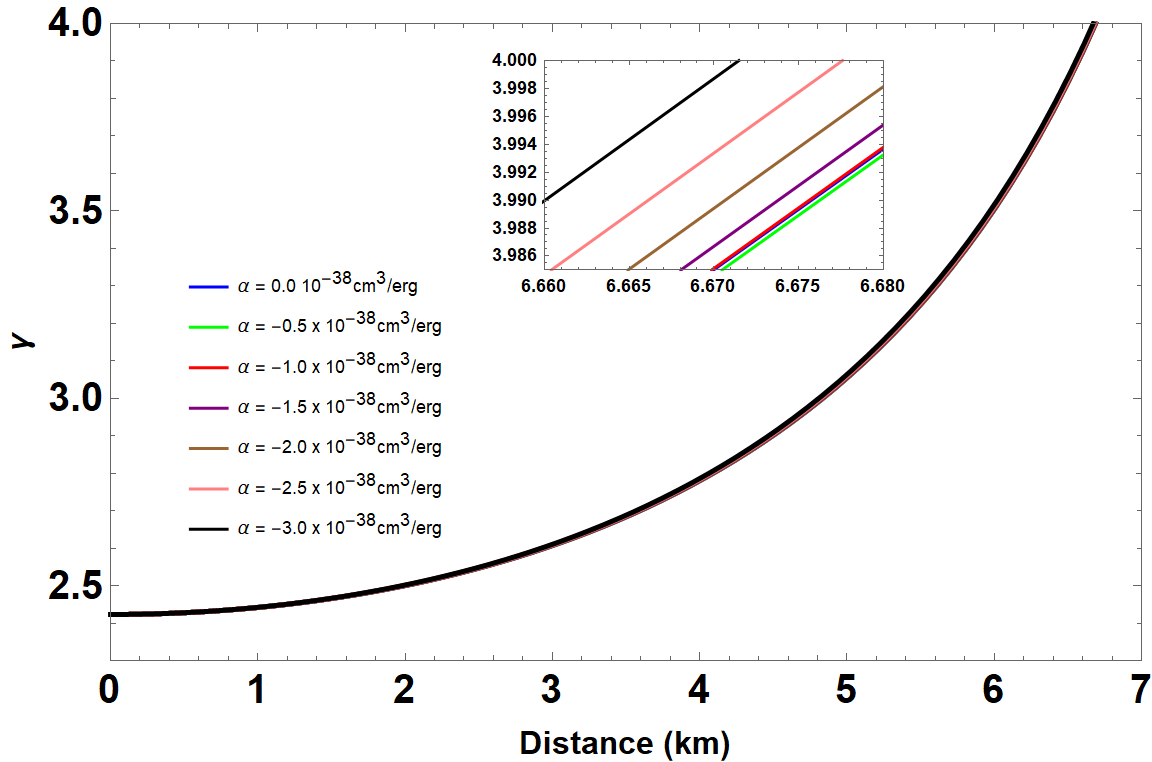}
    \includegraphics[width = 7.5 cm]{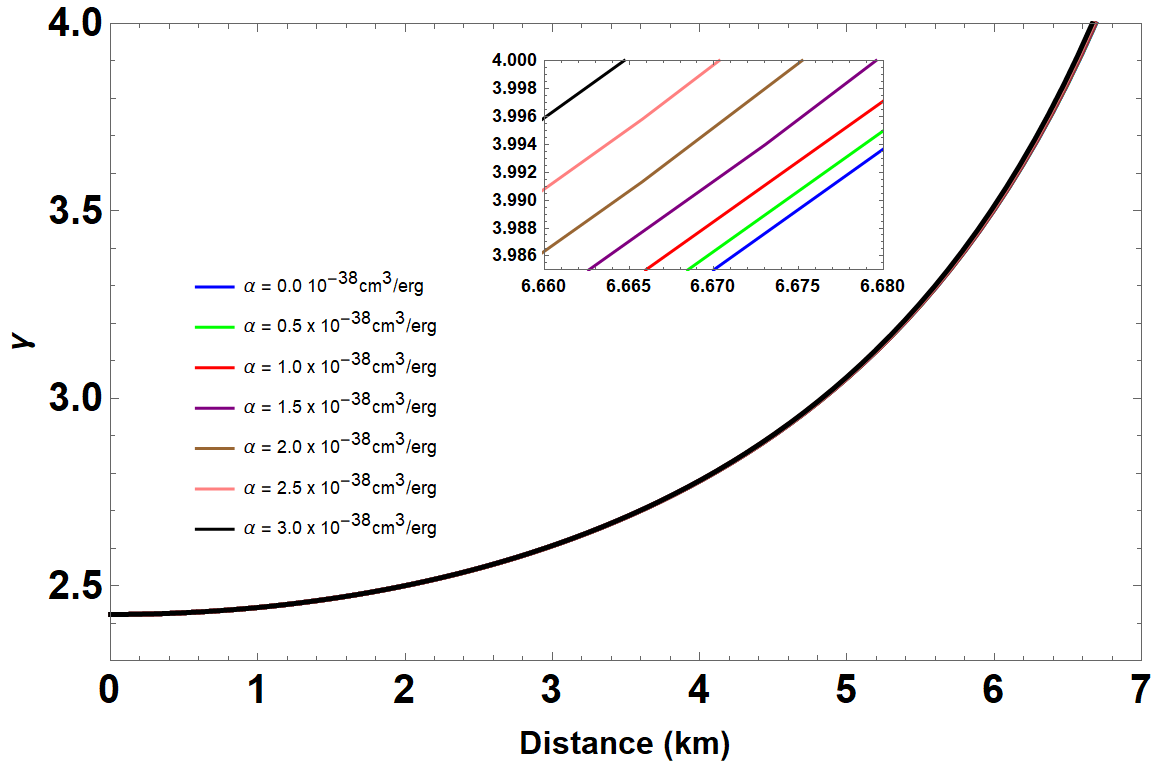}
    \includegraphics[width = 7.5cm]{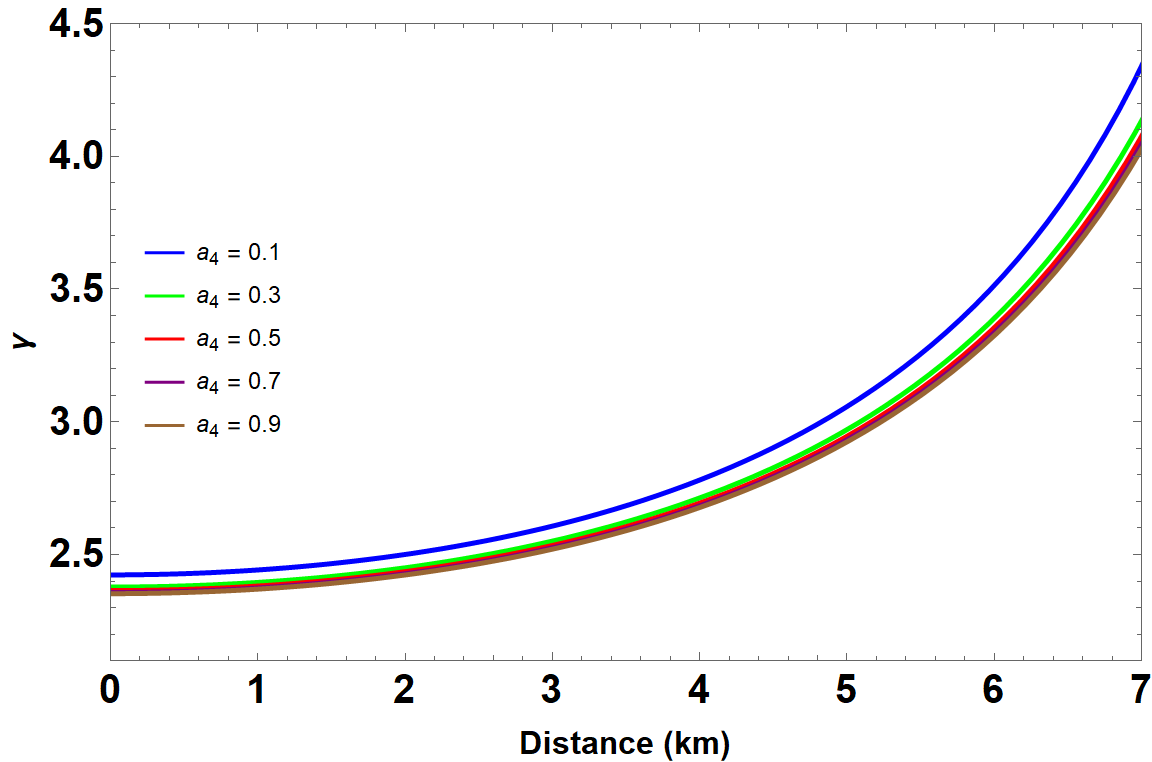}
    \caption{The figure demonstrates the adiabatic index, $\gamma$, with the same set of parameters, see Figs. \ref{figVaryAlpha}, \ref{figVaryAlpha2_positive} and  \ref{figVaryA4}, respectively.}
    \label{figGammaVaryA4}
\end{figure}

For the case of relativistic polytropes, the value of $\gamma$ should be $\gamma> \gamma_{cr} = 4/3$ \cite{Glass}, where $\gamma_{cr}$ is the critical adiabatic index. This value is strictly dependent on the ratio $\rho/P$ at the center of the star. To check this criterion for interacting quark EoS, we plot $\gamma$
as a function radial coordinate in Fig. \ref{figGammaVaryA4} for several representative values of $\alpha$ and $a_4$. Similar to static fluid spheres in GR,
we have seen that the criterion $\gamma > \gamma_c$, ensures the dynamical stability, and
the configuration will prevent progressive collapse.

\section{ concluding remarks }\label{sec5}
Quick development in the astronomical observable techniques provides a strong constraint on the high-density region of the nuclear matter EoS, and consequently, the inner composition of NSs.
These indications might be a possible appearance of quark matter in compact stars, which may be the true ground state of strong interaction. In this paper, we explore the possibility and the corresponding implications of the idea that compact stars are made of interacting quark EoS (\ref{Prad1}), which strictly depends on the interaction parameter, $a_4$ and the bag constant, $B$.

We consider QS models in the framework of energy-momentum squared gravity (EMSG). Specifically, EMSG is treated as a simple generalization to Einstein's general relativity by adding a nonlinear $T_{\mu \nu}T^{\mu \nu}$ term to the generic action.
The modified TOV equations are solved numerically to obtain the mass-radius ($M-R$) relations for the interacting quark EoS. We have studied the stability of QS and their physical properties
depending on the values of coupling constant $\alpha$ and the interaction parameter $a_4$. As a first step in our study, we consider the negative values of $\alpha$ and plot the $M-R$ relations and the compactness ($M/R)$ of QSs in EMSG in Fig. \ref{figVaryAlpha}. The star attains its maximum mass for decreasing values of $\alpha$. We note that EMSG leads to larger masses for QSs compared with GR (see Table \ref{tableVaryAlpha}). At the same time the maximum compactness decreases with decreasing values of $\alpha$.  Next we explore the role of positive values of $\alpha$. We see that the maximum mass decreases for increasing values of $\alpha$ and Fig. \ref{figVaryAlpha2_positive} shows some associated $M-R$ relations. This is interesting in the sense that, this EoS leads to maximum mass 1.782 $M_{\odot}$ when $\alpha =0$ i.e., 
GR solution. Table \ref{tableVaryAlpha2_pos} summarizes the compactness for $\alpha > 0$. One sees that the maximum compactness decreases with increasing the values of $\alpha$. However, it is found that compactness stays nearly the same for both cases. Finally, we  vary the interacting parameter $a_4$ and obtained the $M-R$ relations in Fig. \ref{figVaryA4}.
We have seen that the increasing values of $a_4$, the maximum masses and their corresponding radii have increased. These models have $M_{\text{max}}$ = 1.759 $M_{\odot}$ and  $\frac{M}{R} \sim 0.2688$ when $a_4 = 0.9$ for a range of plausible values of $\alpha$.

To check the stability of QSs constructed in EMSG, we study the static stability criterion and the adiabatic index $\gamma$, which indicate the onset of instability. To distinguish the regions made by stable and unstable configurations, we infer the condition
$\frac{\partial M(\rho_c)}{\partial \rho_c}$ $\lessgtr 0$. It is seen from the Fig. \ref{fig_ME} that the configurations on the segments $\frac{\partial M(\rho_c)}{\partial \rho_c}> 0$ are always stable against radial oscillations but does not guarantee stellar stability in EMSG. Finally, we investigate the critical values of the adiabatic index, and our results clearly show that $\gamma> \gamma_{cr} = 4/3$ (see Fig. \ref{figGammaVaryA4}), which means a stable QS may exist in EMSG theory.

\begin{acknowledgments}
 T. Tangphati was supported by King Mongkut's University of Technology Thonburi's Post-doctoral Fellowship. A. Pradhan thanks to IUCCA, Pune, India for providing facilities under associateship programmes.
\end{acknowledgments}\

\end{document}